\documentclass[prl,aps,twocolumn,floatfix,showpacs]{revtex4}
\usepackage{graphicx,graphics,psfrag,amsmath,calc}
\usepackage{epsfig}
\usepackage{color}

\begin{document}

\title{
Who is the Lord of the Rings: Majorana, Dirac or
Lifshitz? \\
The Spin-Orbit-Zeeman Saga in Ultra-cold Fermions. 
}

\author{
Kangjun Seo, Li Han and C. A. R. S{\'a} de Melo
}

\affiliation{
School of Physics, Georgia Institute of Technology, Atlanta, Georgia 30332, USA
}

\date{\today}

\begin{abstract}
We discuss the emergence of rings of zero-energy excitations in momentum
space for superfluid phases of ultra-cold fermions when spin-orbit, Zeeman
fields and interactions are varied. We show that phases containing
rings of nodes possess non-trivial topological invariants, and that 
phase transitions between distinct topological phases belong to the
Lifshitz class. Upon crossing phase boundaries, existing massless 
Dirac fermions in the gapless phase anihilate to produce 
bulk zero-mode Majorana fermions at phase boundaries and 
then become massive Dirac fermions in the gapped phase.
We characterize these tunable topological phase transitions 
via several spectroscopic properties, including excitation spectrum, 
spectral function and momentum distribution. Since the emergence 
or disappearance of rings leads to topological transitions 
in momentum space, we conclude that Lifshitz is the lord of the rings.
\end{abstract}
\maketitle

%
%

Ultra-cold atoms have now become standard laboratories
to test for existing or new theoretical 
ideas in atomic, condensed matter, nuclear and astrophysics. 
The major appeal found in these table-top experiments is the
ability to tune interactions, populations, species of atoms and 
dimensionality - which constitute the standard toolbox
for investigations of interacting bosonic or fermionic systems.
Very recently a new tool has been added to the toolbox:
the ability to tune simultaneously spin-orbit and Zeeman fields
in a system of ultra-cold bosonic atoms~\cite{spielman-2011}. 
The same tool can also be used to study ultra-cold 
fermionic atoms~\cite{spielman-2011, sinova-2009, chapman-sademelo-2011}
and to simulate different condensed matter systems such as 
topological insulators~\cite{kane-2005}, 
non-centrosymmetric superconductors~\cite{gorkov-2001} and
non-equilibrium systems~\cite{galitski-2007}, where
spin-orbit coupling of the Rashba-type~\cite{rashba-1960} 
is encountered.

This direct connection to condensed matter physics inspired 
a new direction in ultra-cold fermionic atoms where spin-orbit
coupling of the Rashba-type has been very recently 
investigated~\cite{shenoy-2011,chuanwei-2011, zhai-2011, hu-2011, 
iskin-2011,han-2011}.
However, spin-orbit fields currently realized in experiments 
involving ultra-cold atoms correspond to an equal superposition of
Rashba~\cite{rashba-1960}
$
{\bf h}_R ({\bf k})
= 
v_R
(
- k_y {\hat {\bf x}} 
+
k_x {\hat {\bf y}}
)
$
and Dresselhaus~\cite{dresselhaus-1955}  
$
{\bf h}_D ({\bf k})
= 
v_D 
(
k_y {\hat {\bf x}} 
+
k_x {\hat {\bf y}}
) 
$
fields, leading to equal-Rashba-Dresselhaus (ERD) 
form~\cite{spielman-2011,han-2011} 
${\bf h}_{ERD} ({\bf k}) = v {k_x} {\hat{\bf y}}$, 
where $v_R = v_D = v/2$. 
Other forms of spin-orbit fields 
require additional lasers and create further experimental 
difficulties~\cite{dalibard-2010}, such that the 
current Zeeman-spin-orbit Hamiltonian created in 
the laboratory is 
\begin{equation}
\label{eqn:zeeman-spin-orbit}
{\bf H}_{ZSO} ({\bf k})
= 
- h_z \sigma_z - h_y \sigma_y - h_{ERD} ({\bf k}) \sigma_y
\end{equation}
for an atom with center-of-mass momentum ${\bf k}$ and 
spin basis $\vert\uparrow \rangle$, 
$\vert\downarrow \rangle$.  
The fields 
$h_z = - \Omega_R/2$,  $h_y = -\delta/2$, 
and $h_{ERD} ({\bf k}) = v k_x$
can be controlled independently
and can be used to explore phase diagrams 
as achieved in $^{87}$Rb experiments~\cite{spielman-2011}.
Here, $\Omega_R$ is the Raman coupling
and $\delta$ is the detuning.

%
%

{\it Hamiltonian:}
To investigate artificial spin-orbit and Zeeman fields 
in ultra-cold Fermi superfluids, we  
start from the Hamiltonian density
$
{\cal H} ({\bf r})
=
{\cal H}_0 ({\bf r})
+
{\cal H}_I ({\bf r}),
$
with $\hbar = 1$. The single-particle contribution is 
\begin{equation}
\label{eqn:hamiltonian-single-particle} 
{\cal H}_0 ({\bf r}) 
=
\sum_{s, s^\prime} 
\psi^{\dagger}_{s} ({\bf r}) 
\left[ 
K (\hat {\bf k} ) {\bf 1} 
+
{\bf H}_{ZSO} (\hat {\bf k})
\right]_{s s^\prime} 
\psi_{s^\prime} ({\bf r}),
\end{equation}
where
$
K (\hat {\bf k}) 
= 
{\hat {\bf k}}^2/(2 m) 
- 
\mu
$ 
is the kinetic energy 
relative to the chemical potential $\mu$, 
$[ {\bf H}_{ZSO} (\hat {\bf k}) ]_{s s^\prime}$ 
are the matrix elements of the Zeeman-spin-orbit 
matrix Hamiltonian defined in Eq.~(\ref{eqn:zeeman-spin-orbit}),
${\hat {\bf k}} = - i \nabla$ is the momentum operator,
and $\psi^\dagger_{s} ({\bf r})$ creates fermions 
with spin $s$ at position ${\bf r}$.
The interaction term
\begin{equation}
\label{eqn:hamiltonian-interaction}
{\cal H}_I ({\bf r})
=
-g
\psi^{\dagger}_{\uparrow} ({\bf r})
\psi^{\dagger}_{\downarrow} ({\bf r})
\psi_{\downarrow} ({\bf r})
\psi_{\uparrow} ({\bf r}),
\end{equation}
is local and $g$ represents a contact interaction strength.
We define the total number of fermions as 
$N = N_\uparrow + N_\downarrow$,
and the induced population imbalance as 
$P_{\rm ind} = (N_\uparrow - N_\downarrow)/N$. 
We choose our scales through 
the Fermi momentum $k_{F}$ defined from the 
total density of fermions
$
n
= 
n_\uparrow 
+ 
n_\downarrow
=
k_{F}^3/(3\pi^2). 
$ 
This choice leads to the Fermi energy 
$\epsilon_{F} = k_{F}^2/2m$ 
and to the Fermi velocity $v_{F} = k_{F}/m$,
as energy and velocity scales respectively.

We focus on the zero-detuning case $\delta = 0$ $(h_y = 0)$, 
use the basis
$
\psi_{\uparrow}^\dagger ({\bf k}) \vert 0\rangle
\equiv \vert {\bf k} \uparrow \rangle,
$ 
$
\psi_{\downarrow}^\dagger ({\bf k}) 
\vert 0 \rangle
\equiv \vert {\bf k} \downarrow \rangle,
$ 
where $\vert 0\rangle$ is the vacuum state, 
and write ${\cal H}_0 ({\bf r})$ as the matrix 
\begin{equation}
\label{eqn:hamiltonian-0-momentum}
{\bf H}_{0} ({\bf k}) 
=
K ({\bf k}) {\bf 1}
- h_z \sigma_z
- h_{ERD} ({\bf k}) \sigma_y.
\end{equation}
The interaction Hamiltonian $\cal H_{I} ({\bf r})$ 
can also be converted into momentum space as
$
{\cal H}_{I} ({\bf q})
= 
-g 
b^\dagger ({\bf q})
b ({\bf q}),
$
where the pair creation operator with center
of mass momentum ${\bf q}$ is 
$
b^\dagger ({\bf q})
=
\sum_{{\bf k}}
\psi^\dagger_{\uparrow} ( {\bf k} + {\bf q}/2 )
\psi^\dagger_{\downarrow} (-{\bf k} + {\bf q}/2 ),  
$
%
%
and $g$ can be expressed in terms of the scattering length through 
$
V/g 
= 
-V m/(4\pi a_s) 
+ 
\sum_{\bf k} 1/(2\epsilon_{\bf k}).
$
%

%
%

{\it Helicity Basis:} The matrix ${\bf H}_0 ({\bf k})$ can be diagonalized 
in the helicity basis 
$
\Phi_{\Uparrow}^\dagger ({\bf k}) \vert 0 \rangle
\equiv \vert {\bf k} \Uparrow \rangle,
$
$
\Phi_{\Downarrow}^\dagger ({\bf k}) \vert 0 \rangle
\equiv \vert {\bf k} \Downarrow \rangle,
$ 
via a momentum-dependent SU(2) rotation.
The helicity spins $\Uparrow$ and $\Downarrow$ are aligned or
antialigned with respect to the effective magnetic field 
$
{\bf h}_{\rm eff} ({\bf k}) 
= 
h_z {\hat {\bf z}}
+
h_{ERD} ({\bf k}) { \hat {\bf y} }.
$
The eigenvalues of 
the Hamiltonian matrix ${\bf H}_0 ({\bf k})$ 
are 
$
\xi_{\Uparrow} ({\bf k})  
= 
K ({\bf k}) - \vert {\bf h}_{\rm eff} ({\bf k})\vert  
$
and
$
\xi_{\Downarrow} ({\bf k})  
= 
K ({\bf k}) + \vert {\bf h}_{\rm eff} ({\bf k})\vert,  
$
where
$
\vert {\bf h}_{\rm eff}({\bf k})\vert  
=
\sqrt{
h_z^2 
+ 
h_{ERD}^2({\bf k}) 
}
$
is the magnitude of the effective magnetic field.
The interaction Hamiltonian ${\cal H}_{I}({\bf q})$ 
can be written in the helicity basis as
$
{\widetilde {\cal H}}_{I} ({\bf q})
=
-g
\sum_{\alpha \beta \gamma \delta}
B^\dagger_{\alpha \beta} ({\bf q})
B_{\gamma \delta} ({\bf q}),
$
where pairing is now described by the operator
\begin{equation}
\label{eqn:pairing-operator-helicity-basis}
B_{\alpha \beta}^\dagger ({\bf q})
=
\sum_{\bf k}
\Lambda_{\alpha \beta}
({\bf k}_1, {\bf k}_2)
\Phi_{\alpha}^\dagger ({\bf k}_1 )
\Phi_{\beta}^\dagger  ({\bf k}_2)
\end{equation}
and its Hermitian conjugate, with momentum indices
${\bf k}_1 = {\bf k} + {\bf q}/2$
and 
${\bf k}_2 = -{\bf k} + {\bf q}/2$.
The matrix 
$
\Lambda_{\alpha \beta}
({\bf k} + {\bf q}/2, -{\bf k} + {\bf q}/2)
$
is directly related to the matrix elements of the momentum
dependent SU(2) rotation 
into the helicity basis,
and reveals that the center of mass momentum 
${\bf k}_1 + {\bf k}_2 = {\bf q}$ and
the relative momentum ${\bf k}_1 - {\bf k}_2 = 2{\bf k}$ 
are coupled and no longer independent.

%
%

{\it Tensor Order Parameter:}
From Eq.~(\ref{eqn:pairing-operator-helicity-basis}) it is
clear that pairing between fermions of momenta 
${\bf k}_1$ and ${\bf k}_2$ can occur within the 
same helicity band (intra-helicity pairing) 
or between two different helicity bands 
(inter-helicity pairing). For pairing at ${\bf q} = 0$,
the order parameter for superfluidity is the tensor 
$
\Delta_{\alpha \beta} ({\bf k})
=
\Delta_0
\Lambda_{\alpha \beta} ({\bf k}, -{\bf k}), 
$
where 
$
\Delta_0
= 
- g 
\sum_{\gamma \delta}
\langle 
B_{\gamma \delta} ({\bf 0})
\rangle,
$
leading to 
components:
$
\Delta_{\Uparrow \Uparrow} ({\bf k})
= 
i\Delta_T ({\bf k}) 
{\rm sgn} 
\left[ 
k_x
\right]
$ 
for helicity projection 
$\lambda = +1$;
$
\Delta_{\Uparrow \Downarrow} ({\bf k}) 
= 
- 
\Delta_S ({\bf k}),
$ 
and 
$
\Delta_{\Downarrow \Uparrow} ({\bf k}) 
= 
\Delta_S ({\bf k}),
$
for helicity projection $\lambda = 0$; and
$
\Delta_{\Downarrow \Downarrow} ({\bf k}) 
= 
- i \Delta_T ({\bf k})
{\rm sgn} 
\left[ 
k_x
\right],
$
for helicity projection $\lambda = -1$.
The amplitudes 
$
\Delta_T ({\bf k}) 
= 
\Delta_0 
\vert h_{ERD} ({\bf k}) \vert
/
\vert {\bf h}_{\rm eff} ({\bf k}) \vert
$
and 
$
\Delta_S ({\bf k}) 
= 
\Delta_0 
h_z
/
\vert {\bf h}_{\rm eff} ({\bf k}) \vert
$
reflect the triplet and singlet components 
of the order parameter in the helicity basis. 
The Bloch-sphere relation
$
\vert \Delta_T ({\bf k}) \vert^2
+
\vert \Delta_S ({\bf k}) \vert^2
=
\vert \Delta_0 \vert^2,
$
shows that the singlet and triplet channels in the
helicity basis are not independent.

%
%

{\it Higher angular momentum pairing:}
In the triplet sector 
$\Delta_{\Uparrow \Uparrow} ({\bf k}) $ and 
$\Delta_{\Downarrow \Downarrow} ({\bf k})$
contain not only $p$-wave, but also $f$-wave and higher
odd partial waves, 
as seen from a {\it multipole} 
expansion of 
$
\vert {\bf h}_{\rm eff} ({\bf k}) \vert^{-1} 
= 
\left[
h_z^2 + h_{ERD}^2 ({\bf k}) 
\right]^{-1/2}
$
for finite $h_z$.
Similarly in the singlet sector
$\Delta_{\Uparrow \Downarrow} ({\bf k}) $ and 
$\Delta_{\Downarrow \Uparrow} ({\bf k})$
contain $s$-wave, $d$-wave and 
higher even partial waves, as long
as the Zeeman field $h_z$ is non-zero. 
Higher angular momentum pairing 
occurs because the local (zero-ranged) interaction in the 
$(\uparrow, \downarrow)$ 
spin basis is transformed into a finite-ranged anisotropic
interaction in the helicity basis $(\Uparrow, \Downarrow)$.

%
%

{\it Excitation Spectrum:} The effective Hamiltonian in the helicity basis 
takes the matrix form
\begin{equation}
\label{eqn:saddle-point-hamiltonian-helicity-basis}
\widetilde{\bf H}_{\rm sp} ({\bf k})
=
\left(
\begin{array}{cccc}
\xi_{\Uparrow}({\bf k}) & 0 & 
\Delta_{\Uparrow \Uparrow} ({\bf k}) & \Delta_{\Uparrow \Downarrow} ({\bf k}) \\
0  & \xi_{\Downarrow}({\bf k})&  
\Delta_{\Downarrow \Uparrow} ({\bf k}) & \Delta_{\Downarrow \Downarrow} ({\bf k}) \\
\Delta_{\Uparrow \Uparrow}^* ({\bf k}) & \Delta_{\Downarrow \Uparrow}^* ({\bf k})& 
- \xi_{\Uparrow}({\bf k}) &  0 \\
\Delta_{\Uparrow \Downarrow}^* ({\bf k}) & \Delta_{\Downarrow\Downarrow}^* ({\bf k}) & 
0  & -\xi_{\Downarrow}({\bf k})
\end{array}
\right),
\end{equation}
which has eigenvalues for the highest quasiparticle band
$$
E_1 ({\bf k}) 
=
\sqrt{
\left(
\xi_{h-} 
-
\sqrt{
\xi_{h+}^2 
+
\vert \Delta_S ({\bf k}) \vert^2
}
\right)^2
+
\vert \Delta_T ({\bf k}) \vert^2
},
$$
and for the lowest-energy quasiparticle band,
$$
E_2 ({\bf k}) 
=
\sqrt{
\left(
\xi_{h_-} 
+
\sqrt{
\xi_{h_+}^2 
+
\vert \Delta_S ({\bf k}) \vert^2
}
\right)^2
+
\vert \Delta_T ({\bf k}) \vert^2
}.
$$
The eigenvalues for quasihole bands are
$
E_3 ({\bf k}) 
= 
-
E_2 ({\bf k}) 
$
and
$
E_4 ({\bf k}) 
= 
-
E_1 ({\bf k}).
$
The term 
$
\xi_{h_-}
= 
\left[ 
\xi_\Uparrow ({\bf k}) 
- 
\xi_\Downarrow ({\bf k})
\right]
/
2
$
is the average energy difference between the
helicity bands 
$
\xi_{h_-} 
=
- \vert {\bf h}_{\rm eff} ({\bf k}) \vert;
$
while the energy 
$
\xi_{h_+} 
= 
\left[ 
\xi_\Uparrow ({\bf k}) 
+ 
\xi_\Downarrow ({\bf k})
\right]
/
2
$
is the averaged energy sum of the helicity
bands 
$
\xi_{h_+} 
= 
K ({\bf k}) = \epsilon_{\bf k} - \mu.
$

Notice that $E_1 ({\bf k}) > E_2 ({\bf k}) \ge 0$,
but that only $E_2 ({\bf k})$
can have zeros (nodal regions) corresponding to the
locus in momentum space satisfying the following 
conditions:
a)
$
\xi_{h_-} 
= 
- \sqrt{
\xi_{h_+}^2 
+
\vert \Delta_S ({\bf k}) \vert^2
},
$
which corresponds physically 
to the equality between the effective magnetic field energy
$\vert {\bf h}_{\rm eff} ({\bf k}) \vert$ 
and the {\it excitation energy} for
the singlet component
$
\sqrt{
\xi_{h_+}^2 
+
\vert \Delta_S ({\bf k}) \vert^2
};
$  
and 
b) 
$
\vert
\Delta_T ({\bf k}) 
\vert
=
0,
$
corresponding to zeros 
of the triplet component of
the order parameter 
in momentum space.

{\it Phase Diagram:}
Since only $E_2 ({\bf k})$ can have zeros, the low
energy physics is dominated by this eigenvalue.
In the ERD case, where $\vert h_{ERD} ({\bf k}) \vert = v \vert k_x \vert$,
zeros of $E_2 ({\bf k})$ can occur when $k_x = 0$, 
leading to the following cases:
(a) two possible lines (rings) of nodes
at 
$
(k_y^2 + k_z^2)/(2m) 
= 
\mu  + \sqrt{ h_z^2 - \vert \Delta_0 \vert^2}
$ 
for the outer ring,
and
$
(k_y^2 + k_z^2)/(2m) 
= 
\mu - \sqrt{ h_z^2 - \vert \Delta_0 \vert^2}
$ 
for the inner ring,
when $h_z^2 - \vert \Delta_0 \vert^2 > 0$;
(b) doubly-degenerate line of nodes at
$
(k_y^2 + k_z^2)/(2m) 
= 
\mu 
$ 
for $\mu > 0$,
doubly-degenerate point nodes
for $\mu = 0$, or no-line of nodes
for $\mu < 0$,
when 
$h_z^2 - \vert \Delta_0 \vert^2 = 0$;
(c) no line of nodes when $ h_z^2 - \vert \Delta_0 \vert^2 < 0$.
In addition, case (a) can be refined into cases
(a2), (a1) and (a0). In case (a2), two rings indeed exist provided that 
$\mu > \sqrt{h_z^2 - \vert \Delta_0 \vert^2}$.
However, the inner ring disappears when 
$\mu = \sqrt{ h_z^2 - \vert \Delta_0 \vert^2}$.
In case (a1), there is only one ring when
$
\vert 
\mu
\vert 
< 
\sqrt{h_z^2 - \vert \Delta_0 \vert^2}.
$
In case (a0), the outer ring disappears at
$\mu = - \sqrt{ h_z^2 - \vert \Delta_0 \vert^2}$,
and for $\mu < - \sqrt{h_z^2 - \vert \Delta_0 \vert^2}$
no rings exist. 

\begin{figure} [htb]
\centering
\epsfig{file=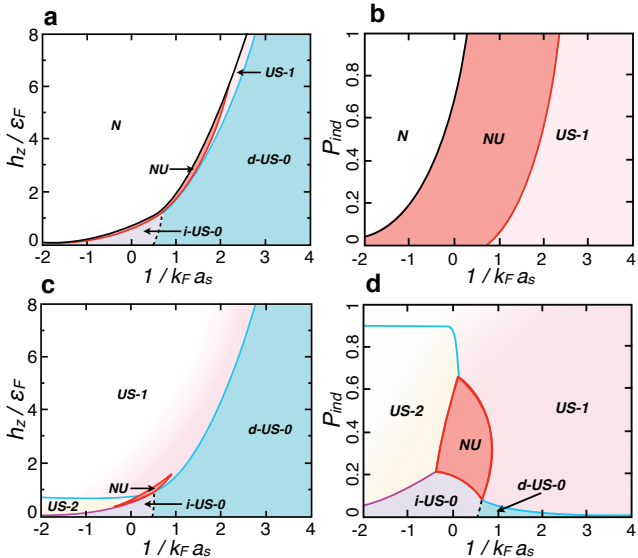,width=1.0 \linewidth} 
\caption{ \label{fig:one} 
(color online) Phase diagrams of $h_z/\epsilon_{F}$ 
and $P_\text{ind}$ versus $1/(k_F a_s)$ for ERD
coupling $v/v_F = 0$ (a), (b) and $v/v_F = 0.28$ (c), (d). 
Uniform superfluid phases are labeled as
US-0 (gapped, either directly or indirectly), US-1 (gapless with one
ring of nodes), and US-2 (gapless with two-rings of nodes).
The NU label describes the region where uniform superfluids are
unstable.
}
\end{figure}

In Fig.~\ref{fig:one}, we show the phase diagrams 
of Zeeman field 
$h_z/\epsilon_{F}$ versus interaction parameter 
$1/(k_F a_s)$ for spin-orbit coupling 
$v/v_F = 0$ (a) and $0.28$ (c), 
as well as induced population imbalance $P_\text{ind}$ 
versus $1/(k_F a_s)$ for $v/v_F = 0$ (b) and $0.28$ (d).
We label the uniform superfluid phases with zero, 
one or two rings of nodes as US-0, US-1, and US-2, respectively.
Non-uniform (NU) phases also emerge in regions where uniform phases 
are thermodynamically unstable. Possible NU phases include
phase separation, modulated superfluids and supersolid.
The US-2/US-1 phase boundary is determined
by the condition $\mu = \sqrt{h_z^2 - \vert \Delta_0 \vert^2}$,
when $\vert h_z \vert > \vert \Delta_0 \vert$;
the US-0/US-2 boundary is determined by
the Clogston-like condition $\vert h_z \vert = \vert \Delta_0 \vert$
when $\mu > 0$, where the gapped US-0 phase  
disappears leading to the gapless US-2 phase;
and the US-0/US-1 phase boundary is determined by 
$\mu = - \sqrt{h_z^2 - \vert \Delta_0 \vert^2}$,
when $\vert h_z \vert > \vert \Delta_0 \vert$.
Furthermore, within the US-0 boundaries, 
a crossover line between an indirectly gapped and a directly gapped 
US-0 phase occurs at $\mu = 0$.

{\it Dirac and Majorana fermions:}
Changes in nodal structures of the order 
parameter are associated with bulk topological phase 
transitions of the Lifshitz class as noted for 
$p$-wave~\cite{volovik-1992, botelho-2005a} and 
$d$-wave~\cite{duncan-2000, botelho-2005b} 
superfluids. 
Such transitions are possible here because 
spin-orbit and Zeeman fields induce higher angular momentum
pairing in the helicity basis.
In the US-1 and US-2 phases near the zeros of 
$E_2 ({\bf k})$, quasiparticles have linear
dispersion and behave as Dirac fermions.
The disappearance of nodal regions (rings) correspond 
to annihilation of Dirac quasiparticles with opposite momenta.
The transition from phase US-2 to indirect-gap i-US-0
occurs through the merger of the two-rings at the phase
boundary followed by the immediate opening of the indirect
gap at finite momentum. However, the transition from 
phase US-2 to US-1 corresponds to the disappearance of 
the inner ring through the origin of momenta, and
the transition from phase US-1 to the direct-gap d-US-0 
corresponds to the disappearance of the last ring also through
the origin of momenta. 

\begin{figure} [htb]
\centering
\epsfig{file=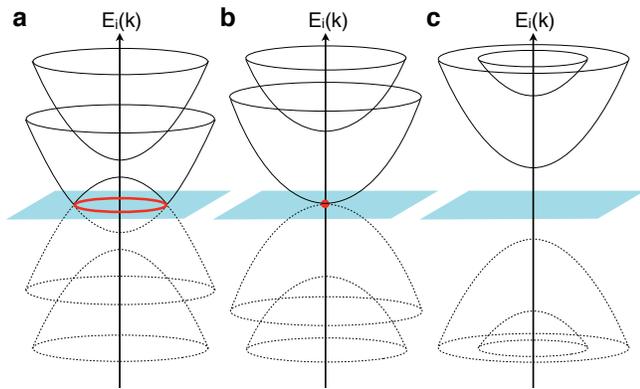,width=1.00 \linewidth} 
\caption{ \label{fig:two} 
(color online) Excitation spectra $E_i ({\bf k})$ in the $(k_x = 0, k_y, k_z)$ 
plane illustrating the Lifshitz transition:
the shrinkage of Dirac rings (US-1 phase) into Majorana 
zero-energy modes (US-1/US-0 phase boundary) and emergence 
of massive Dirac fermions (direct-gap d-US-0 phase).
}
\end{figure}

The last two phase transitions are special because 
the zero-momentum quasiparticles at these phase boundaries 
correspond to Majorana zero 
energy modes if the phase $\varphi ({\bf k})$ of the
spin-orbit field 
$
h_{ERD} ({\bf k}) 
= 
\vert h_{ERD} ({\bf k}) \vert
e^{i \varphi ({\bf k})},
$ 
where 
$
\varphi ({\bf k}) 
= 
{\rm sgn} 
\left[ 
k_x 
\right] 
\pi/2
$
and the phase $\theta ({\bf k})$ 
of the order parameter 
$
\Delta_0 
= \vert \Delta_0 \vert 
e^{i \theta ({\bf k})}
$
hold the relation at zero momentum:
$\varphi ({\bf 0}) = - \theta ({\bf 0})$ $[{\rm mod}(2\pi)]$. 
This can be seen from
an analysis of the eigenfunctions 
$$
\Phi_i ({\bf k}) 
= 
U_{i1,{\bf k}} \psi_{{\bf k} \uparrow} 
+
U_{i2,{\bf k}} \psi_{{\bf k} \downarrow} 
+
U_{i3,{\bf k}} \psi^\dagger_{-{\bf k} \uparrow} 
+
U_{i4,{\bf k}} \psi^\dagger_{-{\bf k} \downarrow} 
$$
corresponding to the eigenvalue $E_i ({\bf k})$ (with $i=2,3$)
where $U_{ij, {\bf k}} = U_{ij} ({\bf k})$ are
the elements of the unitary matrix ${\bf U}$ that 
diagonalizes the Hamiltonian ${\widetilde {\bf H}}_{\rm sp}$.
The emergence of zero-energy Majorana fermions 
requires the quasiparticle (quasihole) to be its own 
anti-quasiparticle (anti-quasihole): 
$
\Phi_i^\dagger ({\bf k}) 
= 
\Phi_i ({\bf k}).
$ 
This happens at zero momentum 
${\bf k} = {\bf 0}$, where the amplitudes 
$U_{i1} ({\bf 0}) = U_{i3}^* ({\bf 0})$ and  
$U_{i2} ({\bf 0}) = U_{i4}^* ({\bf 0})$,
leading to the conditions $\mu^2 =  h_z^2 - \vert \Delta_0 \vert^2$,
and $\varphi ({\bf 0}) = - \theta ({\bf 0})$ $[{\rm mod}(2\pi)]$, 
showing that Majorana fermions exist only at 
the US-1/US-0 and US-2/US-1 phase boundaries. 
In Fig.~\ref{fig:two}, we show the Lifshitz transition from 
US-1 to US-0 phase, where nodal (massless) Dirac Fermions 
in the US-1 phase become bulk zero-mode Majorana fermions at 
the US-1/US-0 phase boundary, and then
massive Dirac fermions in the US-0 phase.

The commonality between bulk Majorana fermions (found here)
and surface Majorana fermions (found in topological insulators
or superconductors) is that both exist at boundaries: 
bulk Majorana zero-energy modes may exist at the phase boundaries
between two topologically distinct superfluid phases, while 
surface Majorana zero-energy modes may exist at the 
spatial (sample) boundaries of a topologically non-trivial 
superconductor.

%
%

{\it Lifshitz Transition:}
The transition between different superfluid phases occurs without 
a change in symmetry of the order parameter tensor 
$\Delta_{\alpha \beta} ({\bf k})$ in the helicity basis,
and thus violates the symmetry-based Landau classification of phase 
transitions. However a finer classification based 
on topological charges can be made via the construction of 
topological invariants~\cite{volovik-1992, nakahara-1990}. 
The number of rings $\ell$ corresponds to the topological charge 
associated with the surfaces of zero-energy quasiparticle excitations. 
Thus, for the US-0 phase $\ell = 0$, while for the US-1 and US-2 phases, 
$\ell = 1$ and $\ell = 2$, respectively. 
%

%
%

\begin{figure} [htb]
\centering
\epsfig{file=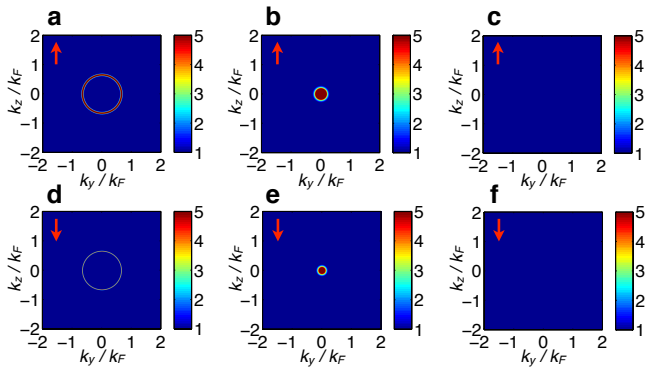,width=1.0 \linewidth} 
\caption{ \label{fig:three} 
(color online) The zero-energy spectral density 
${\cal A}_{s} (\omega = 0, k_x = 0, k_y, k_z)$ 
at $1/(k_F a_s) = 1.0$ and $v/v_F = 0.28$ 
is shown in $(a)$ and $(d)$ for the US-1 phase 
with $h_z/\epsilon_{F} = 1.75$,
in $(b)$ and $(e)$ for the US-1/US-0 phase
boundary with $h_z/\epsilon_{F} = 1.59$,
and in $(c)$ and $(f)$ for the direct-gap 
d-US-0 phase with $h_z/\epsilon_{F} = 1.44$.
}
\end{figure}

{\it Spectral Function:}
An important measurable quantity 
is the single-particle spectral density~\cite{jin-2011}
$
{\cal A}_{s} (\omega, {\bf k}) 
= 
-(1/\pi) 
{\rm Im} G_{s s} (i\omega = \omega + i\delta, {\bf k})
$
for spin $s = \uparrow, \downarrow$, which can be extracted
from the diagonal elements of the matrix 
\begin{equation}
\label{eqn:greens-function}
{\bf G} (i\omega, {\bf k})
=
A (i\omega, {\bf k}) {\bf I}
+ 
{\bf B}(i\omega, {\bf k}) 
\cdot {\bf \sigma},
\end{equation}
where the scalar function is
$
A(i\omega,{\bf k}) 
= 
\{
[
i\omega + K({\bf k})
]
[
(i\omega)^2 - K^2({\bf k}) - \vert \Delta_0 \vert^2 
] 
- \vert h_\text{eff}({\bf k}) \vert^2 
[
i\omega - K({\bf k})
]
\}
/
D(i\omega, {\bf k}),
$
and the vector function components are 
$
B_{x,y}(i\omega,{\bf k}) 
= 
\{
h_{x,y}({\bf k}) 
[ 
\vert h_\text{eff}({\bf k}) \vert^2 
+ 
\vert \Delta_0 \vert^2 
- (i\omega + K({\bf k}))^2 
]
\}
/
D(i\omega, {\bf k})$,
for the transverse and
$
B_{z}(i\omega,{\bf k}) 
= 
h_{z} 
\{
[ 
\vert h_\text{eff}({\bf k}) \vert^2 
- \vert \Delta_0 \vert^2 
- (i\omega + K({\bf k}))^2 
]
\}
/D(i\omega,{\bf k})$ 
for the longitudinal parts.
Here, 
$D(i\omega, {\bf k}) = \prod_{j=1}^{4} (i\omega-E_{j}({\bf k}))$
and $\sigma$ is the vector Pauli matrix.
In Fig.~\ref{fig:three}, we show 
${\cal A}_{s} (\omega, {\bf k})$ 
in the plane of momenta $k_y$-$k_z$ with $k_x = 0$ and 
$\omega = 0$ revealing the existence
of rings of zero-energy excitations in the US-1 and US-2 phases.
In the left panels the spectral densities for the US-1 phase
are shown for spin $\uparrow$ (a),
where the ring is brighter than the ring for 
spin $\downarrow$ (d).
In the middle panels $(b)$ and $(e)$ 
${\cal A}_{s} (\omega, {\bf k})$ at the US-1/US-0
phase boundary is shown revealing the Majorana zero-energy mode.
In the right panels $(c)$ and $(f)$ the spectral densities 
for the US-0 phase vanish at $\omega = 0$, 
since this phase is fully gapped.

%

%
%

\begin{figure} [htb]
\centering
\epsfig{file=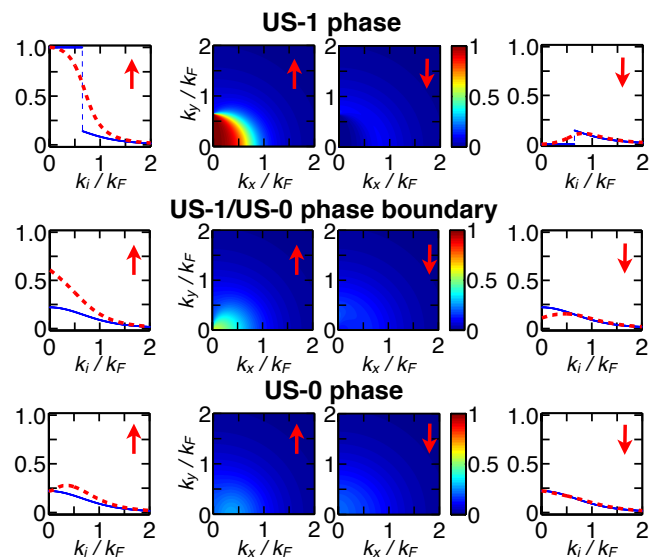,width=1.0 \linewidth} 
\caption{ \label{fig:four} 
(color online) Momentum distribution $n_s ({\bf k})$
for $1/(k_F a_s) = 1.0$ and $v/v_{F} = 0.28$
at the US-1 phase with
$h_z/\epsilon_{F} = 1.75$ (top panels),
at the US-1/US-0 phase boundary with
$h_z/\epsilon_{F} = 1.59$ (middle panels),
at the US-0 phase with 
$h_z/\epsilon_{F} = 1.44$ (lower panels).
The left-most (right-most) panels show the momentum distribution
for spin $\uparrow$ $(\downarrow)$ at $k_z = 0$ versus
$k_y$ (solid-blue line) and versus $k_x$ (dotted-red line).
}
\end{figure}

{\it Momentum Distribution:}
A spectroscopic quantity that is routinely 
measured is the momentum 
distribution
\begin{equation}
n_{s} ({\bf k}) 
=
T \sum_{i\omega} 
\left[ 
A (i\omega, {\bf k}) \pm  B_z (i\omega, {\bf k}) 
\right],
\end{equation}
where the $+$($-$) sign corresponds to spin $\uparrow$ 
$(\downarrow)$. Since $n_{s} ({\bf k})$ depends only on the
energy spectrum $E_j ({\bf k})$ and its derivatives, it is 
an even function of momentum ${\bf k}$. 
In Fig.~\ref{fig:four}, we show $n_s ({\bf k})$
for $1/(k_F a_s) = 1.0$ and $v/v_{F} = 0.28$ at the US-1 phase with
$h_z/\epsilon_{F} = 1.75$ (top panels),
at the US-1/US-0 phase boundary with
$h_z/\epsilon_{F} = 1.59$ (middle panels),
at the US-0 phase with 
$h_z/\epsilon_{F} = 1.44$ (lower panels).
The left-most (right-most) panels show the momentum distribution
for spin $\uparrow$ $(\downarrow)$ at $k_z = 0$ versus
$k_y$ (solid-blue line) and versus $k_x$ (dotted-red line).
It is very important to note the discontinuity of 
$n_{s} ({\bf k})$ at the location of the ring 
of zero-energy excitations in the US-1 phase (top panels), 
its change in behavior as bulk Majorana fermions emerge 
at ${\bf k} = 0$ (middle panels), and the transition to a direct-gap
d-US-0 phase (bottom panels), where the ring of nodes has 
disappeared leading to smooth momentum
distributions $n_{s} ({\bf k})$.  
%

%
%
{\it Summary:}
We showed that the presence of simultaneous Zeeman and
spin-orbit fields induces higher angular momentum pairing
in the helicity basis, and identified topological phase transitions 
of the Lifshitz class via the existence of: rings of nodes in the 
excitation spectra, Dirac quasiparticles, bulk Majorana zero-energy modes, 
and topological charges. Lastly, we characterized 
different topological phases via experimentally measurable quantities such as 
the spectral function and momentum distribution, and 
concluded that Lifshitz is the lord of the rings.

\acknowledgements{We thank ARO (W911NF-09-1-0220) for support.}


\begin{thebibliography}{99}

%
\bibitem{spielman-2011}
Y.-J. Lin, K. Jimenez-Garcia, and I. B. Spielman, 
Nature {\bf 471}, 83 (2011).

%
\bibitem{sinova-2009}
X.-J. Liu, M. F. Borunda, X. Liu, and J. Sinova, 
Phys. Rev. Lett. {\bf 102}, 046402 (2009).

%
\bibitem{chapman-sademelo-2011}
M. Chapman, and C. S{\'a} de Melo,
Nature {\bf 471}, 41 (2011).

%
\bibitem{kane-2005}
C. L. Kane and E. J. Mele,
Phys. Rev. Lett. {\bf 95}, 146802 (2005).

%
\bibitem{gorkov-2001}
L. P. Gor'kov and E. I. Rashba,
Phys. Rev. Lett. {\bf 87}, 037004 (2001).

%
\bibitem{galitski-2007}
T. D. Stanescu, C. Zhang, and V. Galitski,
Phys. Rev. Lett. {\bf 99} 110403 (2007).

%
\bibitem{rashba-1960}
E. I. Rashba,
Sov. Phys. Solid State {\bf 2}, 1109 (1960).

%
\bibitem{shenoy-2011}
J. P. Vyasanakere, S. Zhang, and V. B. Shenoy, 
Phys. Rev. B {\bf 84}, 014512 (2011).

%
\bibitem{chuanwei-2011}
M. Gong, S. Tewari, and C. Zhang,
Phys. Rev. Lett. {\bf 107}, 195303 (2011).

%
\bibitem{zhai-2011}
Z.-Q. Yu, and H. Zhai,
Phys. Rev. Lett. {\bf 107}, 195305 (2011).

%
\bibitem{hu-2011}
H. Hu, L. Jiang, X.-J. Liu, and H. Pu, 
Phys. Rev. Lett. {\bf 107}, 195304 (2011).

%
\bibitem{iskin-2011}
M. Iskin and A. L. Subasi, 
Phys. Rev. Lett. {\bf 107}, 050402 (2011).

%
\bibitem{han-2011}
L. Han and C. A. R. S\'a de Melo, 
arXiv:1106.3613v1 (2011).

%
\bibitem{dresselhaus-1955}
G. Dresselhaus, 
Phys. Rev. {\bf 100}, 580 (1955).

%
\bibitem{dalibard-2010}
J. Dalibard, F. Gerbier, G. Juzeli\ifmmode \bar{u}\else \={u}\fi{}nas, and P. \"Ohberg,
Rev. Mod. Phys. {\bf 83}, 1523 (2011).

%
\bibitem{volovik-1992}
G. E. Volovik, 
{\it Exotic Properties of Superfluid $^3$He},
World Scientific, Singapore (1992).

%
\bibitem{botelho-2005a}
S. S. Botelho and C. A. R. S\'a de Melo,
J. Low Temp. Phys. {\bf 140}, 409  (2005).

%
\bibitem{duncan-2000}
R. D. Duncan, and C. A. R. S\'a de Melo, 
Phys. Rev. B {\bf 62}, 9675 (2000).

%
\bibitem{botelho-2005b}
S. S. Botelho and C. A. R. S\'a de Melo, 
Phys. Rev. B {\bf 71}, 134507 (2005).


%
\bibitem{nakahara-1990}
M. Nakahara, 
{\it Geometry, Topology and Physics},
Adam Hilger, Bristol (1990).

%
\bibitem{jin-2011}
J. P. Gaebler, J. T. Stewart, T. E. Drake, D. S. Jin, A. Perali, P. Pieri	and G. C. Strinati,
Nat. Phys. {\bf 6}, 569 (2010).


\end{thebibliography}
\end{document}